\documentclass[jkps,preprint,fleqn,showpacs,showkeys]{revtex4}
\usepackage{graphicx}
\usepackage{amssymb}
\usepackage{amsmath}
\usepackage{bm}
\begin{document}
\setcounter{page}{0}
\setcounter{page}{0}
\title[]{Consistent theory for causal non-locality beyond Born's
rule}
\author{Wonmin \surname{Son}}
\email{sonwm@physics.org}
\thanks{Fax: +82-2-554-1643}
\affiliation{Department of Physics, Sogang University, Mapo-gu,
Shinsu-dong, Seoul 121-742, Korea}

\date[]{Received 6 August 2007}

\begin{abstract}
According to the theory of relativity and causality, a special
type of correlation beyond quantum mechanics is
possible in principle under the name of {\it non-local box}. The
concept has been introduced from the principle
of non-locality which satisfies relativistic causality. In this
paper, we show that a correlation leading to
the non-local box is possible to be derived consistently if we
release the one of major axioms in quantum mechanics,
{\it Born's rule}. This allows us to obtain a theory which in
one end of the spectrum agrees with the classical probability
and in the other end, agrees with the theory of non-local
causality. At the same time, we argue that the correlation
lies in a space with special mathematical constraints such that
a physical realization of the correlation through
a probability measure is not possible in one direction of its
limit and is possible in the other limit.
\end{abstract}



\maketitle

\section{INTRODUCTION}

Quantum theory predicts a special type of correlation which
allows an immediate action to take place on the state of a
system at a distance \cite{Einstein35}. Due to this special
feature, the existence of extra ordinary correlations can be
taken as a signature indicating whether a system behaves under
the laws of quantum physics \cite{Bell}. Even after extensive
studies, the physical origin of the quantum correlation has not
been unravelled. Specifically, the harmonious co-existence of
this non-local quantum correlation with special relativity has
been taken as the most challenging problem from the inception of
the theories \cite{Aharonov}.

In quantum mechanics, there are axioms that lead us to a
complete description of the theory \cite{Hardy01}. Among them,
{\it Born's rule} gives the probability that a measurement on a
quantum system yields a particular result. The rule is named
after Max Born, who interpreted the wave function of a state as
a probability density which has become one of the key principles
in quantum mechanics \cite{Born26}. It provides a link between
the mathematical formalism of quantum theory and the
experimental realization of quantum measurement. The rule is
responsible for practically all predictions of quantum physics.
The statement of the rule is that if an observable $\hat{X}$
with eigenstates $\{|x_i\rangle\}$ is measured on a system
described by a pure state $|\psi\rangle$, the probability that
the measurement will yield the value $x_i$ is given by
\begin{equation}
\label{eq:born}
p(x_i) =|\langle x_i|\psi\rangle|^2.
\end{equation}
where $p(x_i)$ is the probability for the event of $x_i$.

Historically, numerous attempts have been made to derive Born's
rule from first principles. In Gleason's theorem
\cite{Gleason57}, Born's rule has been formulated from the basic
mathematical assumptions for the probabilities of events as
stated in Eq. (\ref{eq:born}). The probability of quantum
mechanics is therefore dictated by the event structure generated
from the propositions governing measurement \cite{Peres93}.
However, the formulation does not necessarily provide
justification about why nature chooses to behave as the rule
describes. Deutsch tried to answer this question in an intuitive
way \cite{Deutsch99}. He used the non-probabilistic axioms of
quantum theory and classical decision theory to argue that the
probabilities of quantum measurement outcomes can be derived as
per Born's rule. The derivation sparked debates about the charge
of circularity \cite{Barnum00} and gave rise to new derivations
from different angles, {\it e.g.} by Zurek \cite{Zurek05}.
However, the consensus remains that the precise place and of
Born's rule among the axioms of quantum mechanics is not yet
fully understood and continues to be questioned \cite{Zurek11}.
Recently, an experimental test was performed on Born's rule
through the exclusion of multi-order correlation \cite{Sinha08,
Sinha10}.

In this article, it is our intension to identify the implication
of Born's rule on correlations in a bipartite system
and show that a violation of quantum correlation can be obtained
without Born's rule.
box \cite{Popescu94}. This is remarkable because lifting Born's
rule offers a way of obtaining a generalized correlation that
goes beyond quantum mechanics.
We find that lifting Born's rule offers a way of obtaining a
generalized correlation that goes beyond quantum mechanics,
reaching a non-local box \cite{Popescu94} as a limiting case.
In our study, Born's rule is removed in such a way as to remain
consistent with special relativity so that causal non-locality
is still satisfied. Our observations allow us to conclude that
without Born's rule, communication complexity can become trivial
thus the theory becomes unphysical. We start our discussion by
explaining the relationship between the theory of relativity and
non-locality.

\section{Correlation function for Bell's inequality} 
In his historical lecture\cite{Aharonov}, Aharonov conjectured
that {\it non-locality} and {\it relativistic causality} are the
two main elements that specify quantum indeterminacy.
Specifically, he argued that the non-local character of a
quantum system can be regulated by special relativity as per the
quantum correlation predicted by Bell \cite{Bell} and
Clauser-Horne-Shimony-Holt (CHSH) \cite{Clauser}.
The CHSH  function is of the form
\begin{equation}
\label{eq:Bell}
{\cal
B}=E(\vec{a},\vec{b})+E(\vec{a},\vec{b}')+E(\vec{a}',\vec{b})-E(\vec{a}',\vec{b}')
\end{equation}
where $E(\vec{a},\vec{b})$ is a correlation function between two
parties. Considering a spin-$1/2$ bipartite system,
$E(\vec{a},\vec{b})$ is defined as the measure of correlation of
spins along the unit vectors $\vec{a}$ and $\vec{b}$. Allocating
the values $+1$ for spin up and $-1$ for spin down the
correlation function can be written as a sum of joint
probabilities
\begin{equation}
E(\vec{a},\vec{b})=p_{\uparrow\uparrow}+p_{\downarrow\downarrow}-p_{\uparrow\downarrow}
-p_{\downarrow\uparrow}=p_{a=b}-p_{a\neq b}
\end{equation}
where $p_{a=b}$ and $p_{a\neq b}$ refer to coincident and
anti-coincident counts, respectively. Using the normalization
condition $p_{a=b}+p_{a\neq b}=1$, the correlation function
becomes,
\begin{equation}
\label{eq:correl2}
E(\vec{a},\vec{b})=2p_{a=b}-1
\end{equation}
and is bounded by $-1\leq E(\vec{a},\vec{b})\leq 1$ because $0
\leq p_{a=b}\leq 1$. Consequently, a simple algebraical
consideration shows us that the function ${\cal B}$ in Eq.
(\ref{eq:Bell}) can take any arbitrary real values up to $4$
without any constraints. However, an actual counting of local
measurement outcomes does not allow the value of ${\cal B}$ to
exceed $2$ \footnote{Under the local realistic model, if one
measures $a=b$, $a=b'$ and $a'=b$, then it is straightforward to
know $a'=b'$. In that case,
$p_{a=b}=p_{a=b'}=p_{a'=b}=p_{a'=b'}=1$ and we know ${\cal
B}=2$.}. In general, $|{\cal B}|\leq 2$. In fact, the local
realistic model imposes a strong constraint on the joint
probabilities given by classical spin systems. For a quantum
mechanically correlated state of a spin-$1/2$ system, the
maximal violation of the inequality goes up to $2\sqrt{2}$,
called the Cirelson bound \cite{Cierlson}. Aharonov conjectured
that the bound is a consequence of special relativity.

However, it turns out that {\it non-locality} is a stronger
notion of quantum statistics than {\it special relativity}
\cite{Popescu94}. It has been shown from the fact that the
correlation which allows the violation of the Bell inequality
over the Cirelson bound, say $|{\cal B}|= 4$, still satisfies
the crucial constraint in the special relativity-{\it nothing
can travel faster than the speed of light}. It implies that
there can be a theory beyond quantum mechanics which satisfies
special relativity.

A physical theory that accounts for a system bounded by the
correlation $2\sqrt{2}\leq |{\cal B}|\leq 4$ has never been
properly formulated. This is partly because there is no known
physical or non-physical theory governing correlations. In the
following sections, we prove inductively that such a theory can
be obtained once Born's rule is discarded.

\section{Quantum mechanical correlation function} 
Based upon Gleason's theorem, a quantum mechanical formalism of
correlations through the local measurements can be constructed
for {\it observables}. The theory for the local measurements of
spin-$1/2$ systems is given by the Pauli spin operators. Local
measurement of a maximally entangled bipartite system in the
singlet state $|\psi^-\rangle$ gives the correlation function,
as follows (see {\it e.g. p} 162 of \cite{Peres93})
\begin{eqnarray}
E_q(\vec{a},\vec{b})&=&\langle \psi^-
|\vec{a}\cdot\vec{\sigma}\otimes
\vec{b}\cdot\vec{\sigma}|\psi^-\rangle\nonumber \\&=&
-\vec{a}\cdot\vec{b}=\cos(\theta_a-\theta_b).
\end{eqnarray}
Comparing it with Eq.(\ref{eq:correl2}), we find
$p_{a=b}=\cos^2{\theta}$ with the parameterization,
$\theta:=|\theta_a-\theta_b|/2$. Considering the underlying
theory of the coincident probability, one can imagine a
coincident probability amplitude $\psi_{a=b}$ which generates
the probability. In this case, Born's rule states that the
square of the absolute value of the amplitude is the probability
of the coincident counts : $p_{a=b}=|\psi_{a=b}|^2$. Due to the
normalization condition $|\psi_{a=b}|^2+|\psi_{a\neq b}|^2=1$,
we arrive at the functional form of the probability amplitudes
\footnote{The precise derivation of the functional form is
provided in the supplementary material.}
\begin{equation}
|\psi_{a=b}|=\cos\theta~~~\mbox{and}~~~ |\psi_{a\neq
b}|=\sin\theta.
\end{equation}
Now, the probability amplitudes are parameterized by a single
non-local parameter $\theta$. Here, we note that the correlation
is a function of local measurement directions $\theta_a$ and
$\theta_b$ only, since we consider the maximally entangled state
to be tested. One can apply this to the CHSH function in
Eq.(\ref{eq:Bell}) and the Cirelson bound of ${\cal
B}=2\sqrt{2}$ is obtained when $(\theta_a, \theta_b,
\theta_{a'}, \theta_{b'})=(0, \pi/4, \pi/2, 3\pi/4)$.

\begin{figure}
\begin{center}
\includegraphics{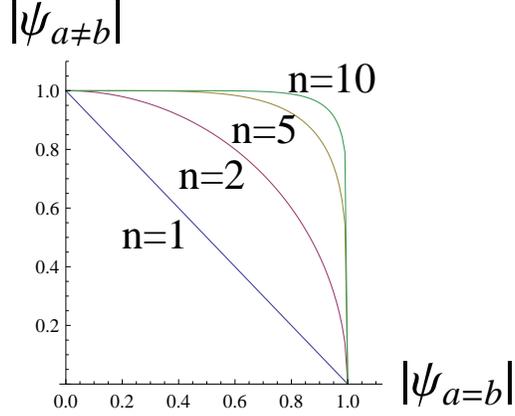}
\end{center}
\caption{The convexity of the probability amplitude under the
normalization constraint $|\psi_{a=b}|^n +|\psi_{a\neq b}|^n=1$.
As $n$ is increased, the function become more convex such that
for a fixed $|\psi_{a=b}|$ the value of $|\psi_{a\neq b}|$
becomes larger as $n$ grows.}
\label{fig:Probability}
\end{figure}

\section{Correlation function without Born's rule} 
Let us consider the consequence of discarding Born's rule. In
general, Eq.(\ref{eq:correl2}) has to hold but the joint
probability is no longer necessarily the absolute square of the
amplitude. So, the correlation function can be written in
general as
\begin{eqnarray}
\label{eq:n-norm correlation}
E_n(\theta_a, \theta_b)&=& |\psi_{a=b}|^n -|\psi_{a\neq
b}|^n\nonumber\\
&=&2 |\psi_{a=b}|^n-1.
\end{eqnarray}
where $\theta_a$ and $\theta_b$ are the local parameters
specifying the local measurements.
The second line uses the normalization condition that
$|\psi_{a=b}|^n +|\psi_{a\neq b}|^n=1$. It also means that the
correlation function can be subject to a single value
parametrization whose physical meaning is directly linked to the
angle between the local measurements at stations A and B. The
probabilities of the coincidence and the anti-coincidence
measurements for different $n$ is plotted for this case in
Fig.\ref{fig:Probability}. The convexity of the function
increases for larger $n$, and becomes a step function in the
limit of $n\rightarrow \infty$.

Motivated by the transformation from Cartesian to polar
coordinates, one can define the angle $\theta$ by $\tan\theta=
|\psi_{a=b}|/|\psi_{a\neq b}|$, $0<\theta<\pi/2$. The
correlation function then becomes $E_n(\theta_a,\theta_b)=1-2
\tan^n\theta/(1+\tan^n\theta)$. When $n=2$, we have
$E_2(\theta)=E_q(\theta)$. The non-local box can be obtained
once $E_{\infty}(\theta_a,\theta_b)$ and is constructed when
$\lim_{n\rightarrow \infty} \tan^n \theta=0$ when $ 0< \theta
<\pi/4$ and $\lim_{n\rightarrow \infty} \tan^n\theta=\infty$
when $\pi/4<\theta <\pi/2$. Generally, any theory producing
correlation $E_{n}(\theta_a,\theta_b)$ with integer $n$, $3\le n
< \infty$, implies the existence of a system which is
asymptotically approaching the non-local box.

However, after a careful inspection, one realizes that the
parametrization $\tan\theta =|\psi_{a=b}|/|\psi_{a\neq b}|$ for
the correlation function $E_{n}(\theta)$ is not consistent with
the local realistic model when $n=1$. Therefore, the
parameterization is not acceptable except for a quantum
mechanical case with $n=2$. The discrepancy occurs due to the
convexity of the tangent function, which means that an increment
of parameter $\theta$ is not uniformly distributed over the
change of the probability amplitudes $|\psi_{a=b}|$ and
$|\psi_{a\neq b}|$. In other words, the parameter does not
produce the uniform distribution of the probability amplitudes
$|\psi_{a=b}|$ and $|\psi_{a\neq b}|$ in the n-norm preserving
space.

To satisfy consistency with a realistic model when $n=1$, one
should find the function $F_n(\theta):=|\psi_{a=b}|$ that
satisfies $|\psi_{a=b}|^n+|\psi_{a\neq b}|^n=1$ together with an
extra condition,
\begin{equation}
\label{eq:uniformity1}
\Big(\partial|\psi_{a=b}|\Big)^2+\Big(\partial|\psi_{a\neq
b}|\Big)^2 \propto \Big(\partial\theta\Big)^2
\end{equation}
which resembles the metric property in geometry. This condition,
namely 2-norm uniformity condition, means that the displacement
of the parameter $\theta$ is uniformly distributed over the
change of mutually exclusive probabilities. After some algebra
\footnote{From the completeness condition, the n-norm uniformity
condition becomes $dF_n/d\theta [1+(1-F_n^n)^{2(1/n-1)}
F_n^{2(n-1)}]^{1/2}= \mbox{const.}$ and the substitution of
$F_n^n=x$ results that $\int dx [x^{2(1/n -1)}
+(1-x)^{2(1/n-1)}]^2= c\theta$ where $c$ is an arbitrary
constant. The constant takes the role of scaling $\theta$. The
equation above implies that $G[x]= c\theta$ where $G[x]:=\int dx
[x^{2(1/n -1)} +(1-x)^{2(1/n-1)}]^2$ and it can be written that
$x= G^{-1}(\theta)= F_n^n(\theta)$.}, one realizes that the
function $F_n(\theta)= [G^{-1}_n(\theta)]^{1/n}$ can be found in
a functional form of integration as
$G_n(x) =\frac{1}{n} \int dx
[x^{-2(1-\frac{1}{n})}+(1-x)^{-2(1-\frac{1}{n})}]^{1/2}.$
Based upon the condition, the functional form of the correlation
and the probability amplitude in Eq.(\ref{eq:n-norm
correlation}) can be obtained as
\begin{equation}
\label{eq:correlation_n}
E_n(\theta)=2 G^{-1}_n(\theta)-1~~~\mbox{and}~~~
|\psi_{a=b}|= G^{-1}_n(\theta)
\end{equation}
which reproduces generic theories from classic and quantum to
non-local box. The correlation regulates the change of the Bell
function in a way that the system never goes beyond the local
realistic model when $n=1$ and reaches a quantum bound when
$n=2$. Analytic expressions of the correlation when $n=1, 2$ are
\begin{equation}
E_1(\theta)=1-\frac{4\theta}{\pi} ~~~\mbox{and}~~~
E_2(\theta)=\cos2\theta
\end{equation}
which coincide with the classical spin system and quantum
mechanics. For the case of {\it classical} spin systems, a
realistic model of the correlation is possible as described by
$E_1(\theta)$ \cite{Peres93} which follows from a realistic spin
system existing in a unit sphere. In this setup, a spin
measurement value is determined by cutting the equatorial plane
of the sphere which is perpendicular to the measurement
direction. The value of spin measurement is $1$ for the spin
pointing in one half of the sphere and is $-1$ for the spin in
the opposite side of the sphere. If a state of two classical
spins is maximally correlated, it can be proven that the
correlation is linearly proportional to the angle between the
measurement directions of the two sides.

\begin{figure}
\begin{center}
\includegraphics{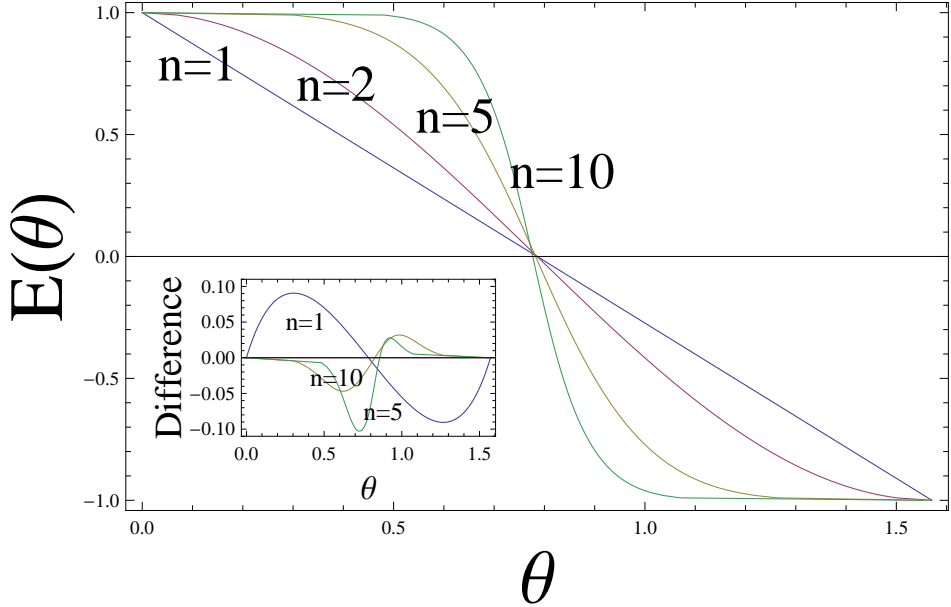}
\end{center}
\caption{Correlation function in Eq.(\ref{eq:correlation_n})
with uniformity condition for various $n$ values. When $n=1$,
the correlation function is a straight line and when $n=2$, the
correlation function becomes a cosine function. As $n$
increases, the correlation function approaches to the step
function.
Inset provides the differences between the correlation functions
with and without the uniformity condition. The difference is
within $10\%$. They coincide only when $n=2$ and $n=\infty$. }
\label{fig:correlation}
\end{figure}

In general, the function $G_n(x)$ with uniformity condition is
not mathematically tractable. After numerical integration, the
correlation with uniformity in Eq.(\ref{eq:correlation_n}) is
plotted in Fig.\ref{fig:correlation}. In fact, the monotonic
behavior of the correlation function coincides asymptotically
with the function without the uniformity condition, although
they are not equivalent. The difference between the correlation
functions is less than $10 \%$ as shown in
Fig.\ref{fig:correlation}. The two correlation functions
coincide only when $n=2$ and $n\rightarrow\infty$. As it
determines the functional form of the correlation uniquely, it
is important to note that the uniformity condition is
nonetheless trivial. The condition imposes a strong constraint
on the correlation and {\it uniquely} determines the functional
form of the correlation with respect to the measurement
parameters.

Consequently, in the limit $n\rightarrow \infty$, the
correlation leads to the non-local box. With a general property
$E_n(\pi/2-\theta)=-E_n(\theta)$, it can be seen that
$E_{\infty}(\theta)=1$ for $0\le \theta\le \pi/4$ and
$E_{\infty}(\theta)=-1$ for $\pi/4\le \theta\le \pi/2$. With
four measurements separated by successive angle $\pi/4$ as
$(\theta_a, \theta_b, \theta_{a'}, \theta_{b'})=(0, \pi/4,
\pi/2, 3\pi/4)$, the CHSH function becomes
\begin{eqnarray}
&&E_{\infty}(\theta_a,\theta_b)+E_{\infty}(\theta_a,\theta_{b'})+E_{\infty}(\theta_{a'},\theta_b)
-E_{\infty}(\theta_{a'},\theta_{b'})
\nonumber\\
&&=3 E_{\infty}(\pi/8)-E_{\infty}(3\pi/8)=4
\end{eqnarray}
which violates the CHSH inequality with the maximal value 4. The
correlation function in the infinite power limit coincides with
the one for the nonlocal box \cite{Popescu94}. Therefore, we can
conclude that the correlation function $E_n(\theta)$ is as
general as to reproduce all the theories, classic, quantum and
nonlocal box consistently.

\section{Relativistic causality and locality}
An important question is whether the arbitrary n-norm preserving
probability theory satisfies the assumption of relativistic
causality. The fact is trivially true due to the normalization
of the conditional probabilities {\it on the marginal
distributions}.
The conditions on the joint probabilities $p(i,j|a,b)$ that the
{\it no-signalling theorem} imposes are $\sum_j
p(i,j|a,b)=p(i|a)$ and $\sum_i p(i,j|a,b)=p(j|b)$
where $p(i|a)$ and $p(j|b)$ are the conditional probabilities of
local measurement $a$ and $b$ with outcomes $i,j\in\{0,1\}$
respectively. It means that the choice of measurement in one
side does not affect the measurement probabilities in the other
side. In quantum theory, the assumption is satisfied due to the
completeness condition of the measurement projection
\footnote{$\sum_j\mbox{Tr}[\hat{P}^A_i\otimes\hat{P}^B_j\rho_{AB}]=
\mbox{Tr}[\hat{P}^A_i\rho_A]$ where $\hat{P}^A_i$ and
$\hat{P}^B_j$ are orthogonal projectors at the sites A and B and
$\rho_A=\mbox{Tr}_B[\rho_{AB}]$.}.

Together with the condition of causality, the standard theory of
probability and Bayesian law for a dichotomic system, one can
derive the no-signalling condition whose joint probability of
the local outcomes is given by
\begin{equation}
\small{p(i,j|a,b)=\frac{p(i|a)+ p(j|b)-1/2}{2}+(-1)^{i+j}
\frac{E_n(a,b)}{4}}
\label{eq:jointprobability}
\end{equation}
where $i,j\in\{0,1\}$ and the correlation function $E_n(\theta)$
in Eq.(\ref{eq:correlation_n}) has been used. The no-signalling
condition always implies that the correlation function can be
written in terms of the local probabilities with local
parameters. (See the supplementary material for an extensive
proof.)

\begin{figure}
\begin{center}
\includegraphics{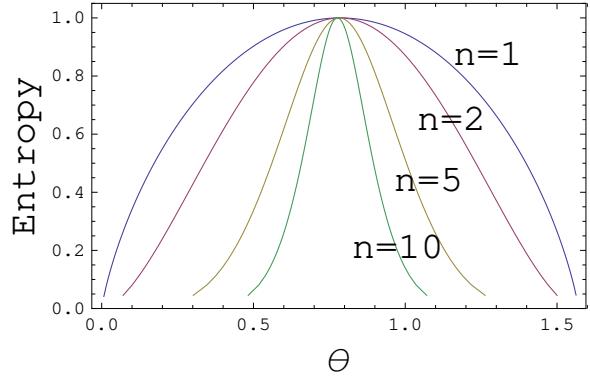}
\end{center}
\caption{Shannon entropy for the maximally correlated state of
the $n$-norm probability theory. The correlation function
(\ref{eq:correlation_n}) and relation
(\ref{eq:jointprobability}) have been used for the calculation
of the entropy. }
\label{fig:entropy}
\end{figure}

Information that is contained in the new statistics can be
represented by Shannon entropy
$S_n(\theta)=-\sum_{ij}p(i,j|a,b)\log p(i,j|a,b)$, plotted in
Fig.\ref{fig:entropy} for a maximally correlated system, {\it
i.e.} $p(i|a)=p(j|b)=1/2$ for $\forall i,j$. In that case, the
convexity of the entropy changes with the measurement angle.
When the spins are measured along the same direction, the system
is completely certain, $S_n(0)=0$, and when the measurements are
orthogonal to each other, the local measurement outcomes of the
two sides are completely random, $S_n(\pi /4)=1$. The entropy
becomes a concave function when $n\ge 2$ from a convex function
when $n=1$. In the limit of $n\rightarrow\infty$, the entropy
becomes a normalized delta function.

In the region of the {\it non-local causal space}, an important
observation has been made through the fundamental principles of
communication complexity \cite{vanDam00}: it has been proven
that any correlation function stronger than quantum mechanics
would render all communication complexity problems {\it trivial}
\cite{vanDam00, Brassard06}. Such an information theoretic
implication has been taken as strong evidence why a correlation
cannot be stronger than quantum mechanics suggests. It means
that $n$ cannot take a value larger than 2 in a physical theory.
On the other hand, it is worth stating that no physical
constraint can be given in the region $1\le n \le2$, even for a
non-integer $n$.

\section{Remark}
{\it Remark -} Born's rule is one of the important axioms in
quantum mechanics that connects most experimental observations
to the theory. However, having discussed the non-local box in
the framework of probability to show the strongest non-local
correlation in a dichotomic bipartite system, we find that
quantum mechanics under Born's rule never achieves such a strong
correlation. In this Letter, we have derived a consistent
correlation function as we discarded Born's rule under the
constraint of the relativistic no-signalling condition. The
correlation function $E_n(\theta)$ in
Eq.(\ref{eq:correlation_n}), is consistent with the local
realistic model when $n=1$ and with quantum non-locality when
$n=2$. When $n>2$, it shows stronger correlations than quantum
non-locality, approaching to the non-local box when
$n\rightarrow\infty$. We note that this is a direct consequence
of releasing Born's rule which renders communication complexity
trivial. Our study assures that Born's rule gives the physically
meaningful correlations.

\begin{acknowledgments}
The authors would like to thank Mr. Kim for helpful discussions
and
gratefully acknowledge support by MOCIE through National R$\&$D
Project for Nano Science and Technology. C. H. C. and W. J. J.
acknowledge the financial support by MOE through BK21
fellowships.
\end{acknowledgments}


\begin{references}
\bibitem{Einstein35} A. Enstein, B. Podolsky and N. Rosen, Phys.
Rev. {\bf 47}, 777 (1935).
\bibitem{Bell} J. S. Bell, Physics {\bf 1}, 195 (1964).
\bibitem{Aharonov} Y. Aharonov, unpublished lecture note; {\it
see also}, Y. Aharonov and D. Rohrlich, Quantum Paradoxes
(WILEY-VCH, 2005).
\bibitem{Hardy01} L. Hardy, arXiv:quant-ph/0101012.
\bibitem{Born26} M. Born, Z. fur Phys., 37, 12 (1926); {\it
English translation}, On the quantum mechanics of collisions, in
Quantum theory and measurement,(ed.) J. A. Wheeler and W. H.
Zurek, (Princeton University Press, 1983).
\bibitem{Gleason57} A. M. Gleason, J. of Math. and Mech., {\bf
6}, 885 (1957).
\bibitem{Peres93} A. Peres, Quantum Theory: Concepts and Methods
(Kluwer academic publishers, London, 1993).
\bibitem{Deutsch99} D. Deutsch, Proc. Roy. Soc. Lond. A, {\bf
455}, 3129 (1999).
\bibitem{Barnum00} H. Barnum, C. M. Caves, J. Finkelstein, C. A.
Fuchs and R. Schack, Proc. Roy. Soc. Lond. A {\bf 456}, 1175
(2000); D. Wallace, Stud. Hist. Phil. Mod. Phys. {\bf 34}, 415
(2003); See also N.P. Landsman, in Compendium of Quantum Physics
(ed.) F.Weinert, K. Hentschel, D.Greenberger and B. Falkenburg
(Springer, 2008).
\bibitem{Zurek05} W. H. Zurek, Phys. Rev. A {\bf 71}, 052105
(2005).
\bibitem{Zurek11} W. H. Zurek, Phys. Rev. Lett. {\bf 106},
250402 (2011).
\bibitem{Sinha08} U. Sinha, C. Couteau, Z. Medendorp, I.
Sollner, R. Laflamme, R. Sorkin and G. Weihs, Proceedings of
Foundations of Probability and Physics 5, Vaxjo, Sweden (2008).
\bibitem{Sinha10} U. Sinha, C. Couteau, T. Jennewein, R.
Laflamme and G. Weihs, Science {\bf 329}, 418 (2010).
\bibitem{Popescu94}S. Popescu and D. Rohrlich, Found. Phys. {\bf
24}, 379 (1994).
\bibitem{Clauser} J. F. Clauser, M.A. Horne, A. Shimony and R.
A. Holt, Phys. Rev. Lett. {\bf 23}, 880 (1969).
\bibitem{Cierlson} B.S. Cirel'son, Lett. Math. Phys. {\bf 4}, 93
(1980).
\bibitem{Barrett05}J. Barrett, N. Linden, S. Massar, S. Pironio,
S. Popescu and D. Roberts, Phys. Rev. A {\bf 71}, 022101 (2005).
\bibitem{vanDam00} W. van Dam, Ph.D. Thesis, University of
Oxford (2000), quant-ph/0501159.
\bibitem{Brassard06}G. Brassard, H. Buhrman, N. Linden, A.
Methot, A. Tapp and F. Unger, Phys. Rev. Lett. {\bf 96}, 250401
(2006).
\end{references}
\end{document}